\documentclass{article}
\usepackage{spconf,amsmath,amssymb,booktabs,url,graphicx}
\usepackage[hidelinks]{hyperref}
\title{Sequence-to-Sequence Refinement for\\ EEND-VC Diarization with a Shared Encoder}
\name{Ruoyu Wang$^{1}$, Jun Du$^{1,*}$, Weixin Zhu$^{2}$, Yannan Wang$^{2}$\thanks{\textsuperscript{*}Corresponding author.}}
\address{$^{1}$University of Science and Technology of China, Hefei, China\\
$^{2}$Ethereal Audio Lab, China\\
wangruoyu@mail.ustc.edu.cn\\}

\begin{document}
\maketitle

\begin{abstract}
End-to-end neural diarization with vector clustering (EEND-VC) separates local segmentation from global speaker assignment, leaving speaker confusion (SC) despite strong segmentation. We propose shared-encoder sequence-to-sequence (S2S) refinement, which reuses DiariZen's self-supervised learning (SSL) representations for speaker-conditioned decoding and prior-constrained assignment. We study frozen-feature refinement with WavLM and structurally pruned w2v-BERT 2.0, and extend the latter to joint training of local segmentation and refinement heads. The heads learn separate layer mixtures while their losses update one acoustic encoder. Across eight domains, frozen-feature refinement reduces macro diarization error rate (DER) from 12.95\% to 12.40\% with WavLM and from 12.35\% to 12.08\% with w2v-BERT 2.0, with relative SC reductions of 17.4\% and 14.6\%. It adds approximately 15\% to system real-time factor. The joint system further lowers macro DER to 11.90\%, improving seven of eight domains over non-joint refinement. Error decomposition separates SC-driven refinement gains from detection-driven joint-system gains.
\end{abstract}


\begin{figure*}[t]
\centering
\includegraphics[width=\textwidth]{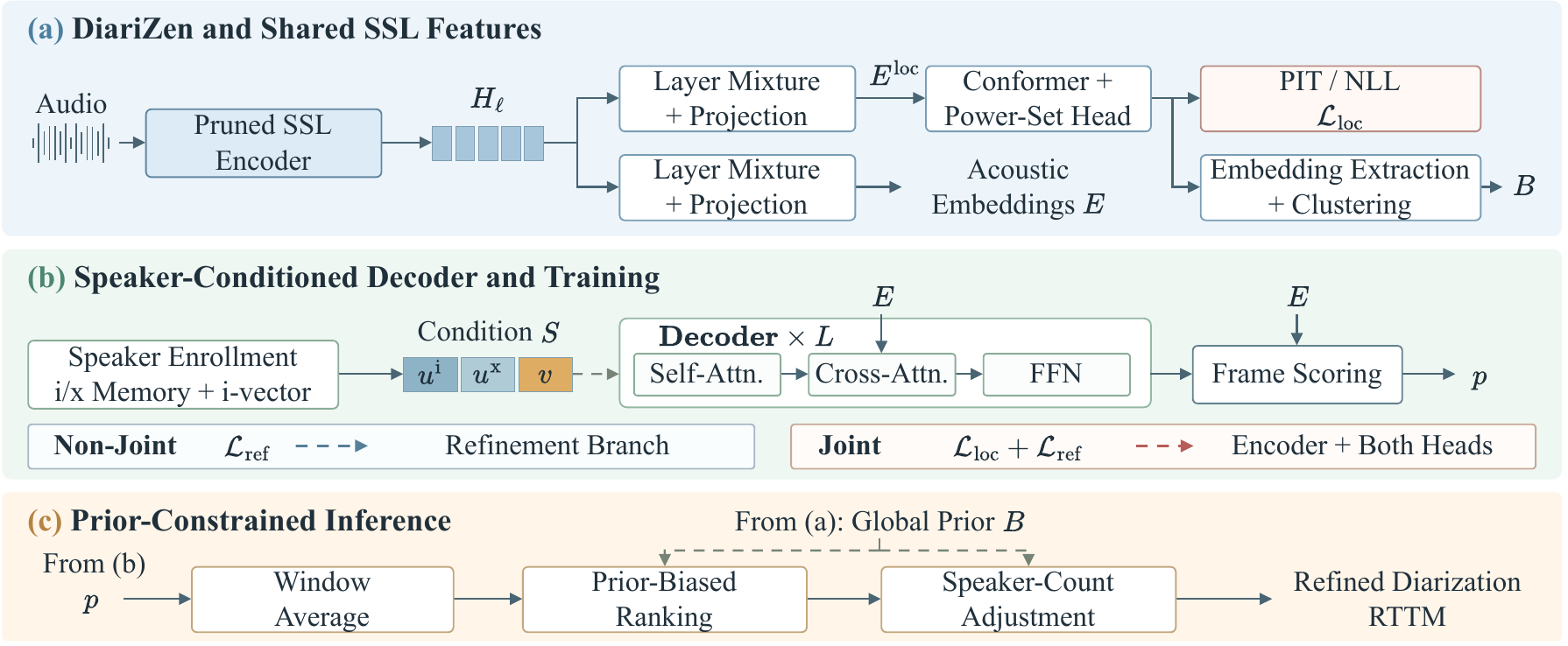}
\caption{DiariZen and shared SSL features (a), speaker-conditioned decoding and training (b), and prior-constrained inference (c). Gray-green dashed arrows denote conditioning; blue/red dashed arrows denote parameter updates.}
\label{fig:s2s_method}
\end{figure*}

\section{Introduction}
\label{sec:introduction}
End-to-end neural diarization (EEND) uses permutation-invariant training (PIT) for frame-level speaker activity prediction~\cite{fujita2019eend}. EEND with vector clustering (EEND-VC) handles long recordings and varying speaker inventories through cross-window clustering~\cite{eendvc2021,palka2026spherevbx}. DiariZen strengthens local segmentation with self-supervised learning (SSL) representations, a Conformer backend, and power-set modeling~\cite{diarizen2025,powerset2023}. Structured pruning improves efficiency alongside its reported state-of-the-art performance across benchmarks~\cite{diarizen2025,pruning2025}. However, segmentation and global identity assignment are optimized separately: correct local activities and speaker counts can still receive incorrect recording-level identities. This mismatch motivates a refinement stage that can reconsider identities while retaining reliable local activity estimates.

Post-processing can revisit these assignments. Neural resegmentation revises local activities~\cite{Bredin_2021}, while target-speaker voice activity detection (TS-VAD) uses enrollment embeddings~\cite{tsvad2020}. Seq2Seq-TSVAD separates acoustic and speaker inputs~\cite{Cheng_2023}; neural speaker diarization using memory-aware multi-speaker embedding with sequence-to-sequence architecture (NSD-MS2S) adds memory-based representations~\cite{yang2024nsdms2s}. Enrollment depends on initial diarization, motivating improved initialization and iterative refinement in USTC's CHiME-8 system~\cite{chime82024}. NTT CHiME-8 initializes NSD-MS2S with EEND-VC outputs but retains a separate acoustic encoder~\cite{ntt2026}. Chunk-level speaker queries also refine DiariZen outputs on DIHARD III~\cite{Tawara_2026}.

Sharing an acoustic encoder can avoid repeated encoding and permit adaptation to both segmentation and refinement. DiariZen's diarization output provides an activity prior for this stage, allowing speaker-conditioned predictions to be constrained by a strong local model. Yet lower diarization error rate (DER) need not imply better speaker assignment: detection-threshold changes can reduce detection errors~\cite{Bredin_2021}. We therefore distinguish speaker confusion (SC) from false alarms (FA) and missed speech (MISS) when evaluating refinement and joint training across domains.

We develop sequence-to-sequence (S2S) refinement by replacing NSD-MS2S's acoustic encoder with DiariZen's SSL representations while retaining its enrollment mechanism and a lightweight decoder. Prior-constrained inference revises identities and overlap counts. Each window undergoes one SSL pass, while independently learned layer mixtures let the two heads use different combinations of SSL representations. We study frozen-feature refinement and shared-encoder joint training of local segmentation and refinement; clustering remains outside the training graph. Eight-domain evaluation with WavLM and pruned w2v-BERT 2.0 shows SC-driven gains from frozen-feature refinement and further detection-driven DER gains from the joint system. Component analysis and identity-correction cases explain these effects.

\section{Method}
\label{sec:method}

\subsection{DiariZen and Shared SSL Features}
DiariZen~\cite{diarizen2025} combines local segmentation and cross-window clustering (Fig.~\ref{fig:s2s_method}(a)). We share a pruned WavLM~\cite{wavlm2022} or w2v-BERT 2.0~\cite{seamless2023} encoder across both heads. Each window undergoes one SSL pass; the two heads independently mix and project its layer outputs:
\begin{equation}
E^{(j)}=P_j\!\left(\sum\nolimits_\ell a_\ell^{(j)}H_\ell\right),
\quad j\in\{\mathrm{loc},\mathrm{ref}\}.
\label{eq:shared_features}
\end{equation}
Here $H_\ell$ denotes an SSL layer output, and $a_\ell^{(j)}$ are unconstrained learned weights. Each $P_j$ applies linear projection and layer normalization (LN), followed by dropout for refinement. We denote the decoder input by $E=E^{(\mathrm{ref})}\in\mathbb R^{T\times d}$.

We apply structured pruning~\cite{pruning2025,w2vpruningsv2025} with hard-concrete gates~\cite{hardconcrete2017} to w2v-BERT 2.0's dual feed-forward modules, attention heads, and convolution channels. Export removes inactive units and folds gate scales into weights; convolution statistics retain the original channel count.

The DiariZen branch maps $E^{(\mathrm{loc})}$ through a Conformer and power-set classifier to anonymous speaker activities. Its local loss $\mathcal L_{\mathrm{loc}}$ uses power-set negative log-likelihood (NLL) with permutation-aligned targets. At inference, these activities guide speaker embedding extraction from audio; clustering across windows produces the global binary prior~$B$.

\subsection{Speaker-Conditioned Decoder and Training}
Figure~\ref{fig:s2s_method}(b) uses acoustic embeddings $E$ in place of the NSD-MS2S acoustic encoder, retaining its i-vector and memory-based enrollment mechanism~\cite{yang2024nsdms2s}. A candidate's non-overlapping prior segments provide an input mask $M_{nt}$ and session i-vector $v_n$~\cite{dehak2011ivector}. Masked averaging of $E$ over $M_{nt}$ gives a query to fixed i-vector and x-vector center banks (i/x memory). Attention across candidates then mixes the retrieved representations. Concatenating the two outputs with $v_n$ gives $s_n=[u_n^{\mathrm i};u_n^{\mathrm x};v_n]$, row $n$ of condition matrix $S$.

The decoder initializes candidate states to zero. Each of $L$ layers injects $s_n$ into self-attention queries/keys and cross-attention queries, while $E$ supplies acoustic keys/values. A feed-forward network (FFN) completes the layer. Final states act as speaker attractors; layer-normalized frame and attractor projections $f_t,g_n$ produce
\begin{equation}
p_{nt}=\sigma\!\left(\frac{\tau}{\sqrt d}f_t^\top g_n+b\right),
\label{eq:frame_scoring}
\end{equation}
with learned positive scale $\tau$ and bias $b$. Independent sigmoids predict all candidates' activities in parallel, allowing overlap without autoregressive decoding.

Both training modes use binary cross-entropy (BCE) for real and sampled negative candidates. Each term averages valid frames; negative targets are zero:
\begin{equation}
\mathcal L_{\mathrm{ref}}=
\overline{\mathrm{BCE}}_{\mathrm{real}}+
\eta\,\overline{\mathrm{BCE}}_{\mathrm{neg}}.
\label{eq:ref_loss}
\end{equation}
\emph{Non-joint training} fixes the SSL encoder and local branch, updating the refinement branch, including its layer mixture and memory query/key projections. It uses paired predicted or oracle priors/profiles with oracle targets.

\emph{Joint training} uses oracle priors/profiles and updates the encoder and both heads with $\mathcal L_{\mathrm{joint}}=\mathcal L_{\mathrm{loc}}+\mathcal L_{\mathrm{ref}}$. Unlike local PIT, refinement supervises enrolled speaker identities.

Both modes exclude overlap from enrollment but retain it in targets. Memory centers and the i-vector extractor stay fixed; clustering, profile extraction, and inference constraints are outside the training graph.

\subsection{Prior-Constrained Inference}
Both modes follow Fig.~\ref{fig:s2s_method}(c). Non-overlapping regions of DiariZen's prior $B$ supply $M$ and session i-vectors; candidates without usable profiles are excluded. During the local pass, the refinement branch caches its acoustic embeddings $E$. Once global tracks and enrollment profiles are available, decoding reads this cache without re-encoding the waveform. Unlike $M$, $B$ retains overlap. Identity-aligned window probabilities are equally averaged to obtain $\bar p_{nt}$.

Prior-biased ranking combines $\bar p$ and $B$:
\begin{equation}
r_{nt}=\operatorname{logit}(\bar p_{nt})+\beta B_{nt},
\qquad k_t=\sum_n B_{nt}.
\label{eq:prior_fusion}
\end{equation}
We first select the top $k_t$ identities. If $k_t>1$ and the weakest selected probability is at most $\theta_{\mathrm{drop}}$, the count decreases by one. Otherwise, an unselected probability at least $\theta_{\mathrm{add}}$ increases a positive count by one. We then reselect identities at the adjusted count. An enrolled candidate inactive in $B$ at a frame can replace an incorrectly assigned speaker if its decoder evidence overcomes the prior bias. Ranking corrects identities at unchanged count; count adjustment can restore missed overlap or remove a spurious speaker. Silent frames remain silent and the last active speaker is retained. The profile-filtered prior's speech support is preserved, while speaker activities and boundaries can change.

\suppressfloats[t]
\setcounter{table}{1}
\begin{table}[t]
\centering
\caption{H20 RTF for frozen-feature refinement (batch 64). DiariZen includes clustering; Refine. is the full increment.}
\label{tab:rtf}
\small
\setlength{\tabcolsep}{2.5pt}
\begin{tabular*}{\linewidth}{@{\extracolsep{\fill}}lcccc@{}}
\toprule
\textbf{Front end} & \textbf{DiariZen} & \textbf{Refine.} & \textbf{Total} & \textbf{Increase} \\
\midrule
WavLM & 0.0177 & 0.0027 & 0.0204 & 15.30\% \\
w2v-BERT 2.0 & 0.0180 & 0.0027 & 0.0207 & 15.03\% \\
\bottomrule
\end{tabular*}
\end{table}

\setcounter{table}{0}
\begin{table*}[!t]
\centering
\setlength{\abovecaptionskip}{4pt}
\caption{DER by dataset and macro error components (\%) under BF16 SSL inference. Datasets are ordered by overlap. DiariZen includes clustering; Joint uses shared-encoder training. Bold/underline mark lowest/second-lowest values per column. DER uses unrounded components.}
\label{tab:main_der}
\small
\setlength{\tabcolsep}{3pt}
\begin{tabular*}{\textwidth}{@{\extracolsep{\fill}}l*{8}{c}|*{4}{c}@{}}
\toprule
& \multicolumn{8}{c|}{\textbf{DER by Dataset}} & \multicolumn{4}{c}{\textbf{Macro Average}} \\
\cmidrule(lr){2-9}\cmidrule(l){10-13}
\textbf{System} & \textbf{RAMC} & \textbf{VoxC} & \textbf{AIS} & \textbf{DH3} & \textbf{MSD} & \textbf{AMI} & \textbf{AliM} & \textbf{NSF} & \textbf{FA} & \textbf{MISS} & \textbf{SC} & \textbf{DER} \\
Overlap (\%) & 0.0 & 3.1 & 4.9 & 9.4 & 12.4 & 14.6 & 20.4 & 30.1 & & & & \\
Speakers & 2 & 1--21 & 5--7 & 1--9 & 2--4 & 3--4 & 2--4 & 3--7 & & & & \\
\midrule
DiariZen (WavLM) & \textbf{10.99} & 9.11 & \underline{9.99} & 14.48 & 15.77 & 13.91 & 10.81 & 18.50 & \underline{3.87} & 5.64 & 3.44 & 12.95 \\
\quad + S2S Refinement & 11.05 & 8.91 & \textbf{9.58} & 14.42 & 15.44 & 13.16 & 10.11 & 16.54 & 4.16 & 5.40 & 2.84 & 12.40 \\
\midrule
DiariZen (w2v-BERT 2.0) & \underline{11.01} & 8.88 & 10.69 & 14.22 & 14.45 & 13.44 & 9.69 & 16.43 & 3.92 & 5.45 & 2.98 & 12.35 \\
\quad + S2S Refinement & 11.17 & \underline{8.70} & 10.16 & 14.33 & \underline{14.35} & \underline{13.02} & \underline{9.36} & \underline{15.53} & 4.14 & \underline{5.39} & \textbf{2.55} & \underline{12.08} \\
\midrule
Joint DiariZen (w2v-BERT 2.0) & 11.02 & 8.73 & 11.02 & \textbf{13.78} & 14.49 & 13.21 & 9.91 & 15.91 & \textbf{3.74} & 5.54 & 2.99 & 12.26 \\
\quad + S2S Refinement & 11.08 & \textbf{8.66} & 10.55 & \underline{13.82} & \textbf{14.27} & \textbf{12.74} & \textbf{9.32} & \textbf{14.74} & 4.05 & \textbf{5.28} & \underline{2.57} & \textbf{11.90} \\
\bottomrule
\end{tabular*}
\end{table*}
\setcounter{table}{2}

\section{Experimental Setup}
\label{sec:setup}

\subsection{Data and Evaluation}
We evaluate channel~0 of AISHELL-4 (AIS)~\cite{aishell4}, AliMeeting (AliM)~\cite{alimeeting}, AMI~\cite{ami}, DIHARD III (DH3)~\cite{dihard3}, MSDWild (MSD)~\cite{msdwild}, NOTSOFAR-1 (NSF)~\cite{notsofar1}, RAMC~\cite{ramc}, and VoxConverse (VoxC)~\cite{voxconverse}: 1,209 recordings, including complete NSF evaluation. NIST md-eval scoring uses zero collar, overlap, and each set's UEM. Macro scores equally weight datasets. Table~\ref{tab:main_der} gives overlap/speech duration ratios within scored regions and reference speaker ranges per recording. RAMC has no annotated overlap.

\subsection{Training and Inference}
Pruned w2v-BERT 2.0 has 115.8M parameters (80.05\% reduction) and 25 hidden-state outputs of width 1024. Local segmentation uses a 256-dimensional projection, four Conformer layers, and 16 power-set classes for four speakers. Refinement uses six 512-dimensional decoder layers with eight attention heads and 1024-dimensional FFNs. We reuse NSD-MS2S's 100-dimensional i-vector extractor and two 128-center memory banks\footnote{\url{https://github.com/liyunlongaaa/NSD-MS2S}} (i/x dimensions: 100/256). We pad training samples to eight candidates with negative profiles ($\eta=0.25$).

Both w2v-BERT 2.0 training modes use 16-s windows from eight corpora, 40\% hard-window sampling, and $8\times$ NVIDIA H20 GPUs. Non-joint training freezes the encoder and samples paired DiariZen/oracle priors and profiles with probabilities 0.7/0.3, using oracle targets. AdamW uses learning rate $2\times10^{-4}$, weight decay 0.01, 8\% warmup, linear decay, and batch size 512. We average the five checkpoints with lowest oracle-prior development loss.

Joint training initializes the encoder from post-pruning distillation, reinitializes both heads, and uses oracle priors/profiles. Encoder/head AdamW learning rates are $2\times10^{-6}/10^{-4}$, with zero weight decay and no schedule. Eight samples per GPU and two accumulation steps give an effective batch of 128. Training stops at epoch 14 with patience five; we average the five checkpoints with lowest joint development loss. Local targets retain the four most active speakers without masking omitted-speaker frames.

Pooled development DER selects refinement parameters and, for joint training, chooses between the best epoch and the five-checkpoint average. Non-joint/joint systems use $\beta=1/0.5$ and add/drop thresholds 0.8/0.2, with no temporal smoothing in refinement. Inference uses 16-s windows, a 1.6-s hop, BF16 SSL computation, and FP32 heads. Table~\ref{tab:rtf} reports frozen-feature real-time factor (RTF; processing time/audio duration).

\section{Results}
\label{sec:results}

\subsection{Diarization Across Domains}
Table~\ref{tab:main_der} compares complete systems. Frozen-feature S2S refinement reduces macro DER from 12.95\% to 12.40\% with WavLM and from 12.35\% to 12.08\% with w2v-BERT 2.0, improving seven and six domains, respectively. Both improve all four meeting sets; high-overlap NSF shows the largest reductions, from 18.50\% to 16.54\% and 16.43\% to 15.53\%. Gains also extend to VoxC and AIS, with overlap below 5\%. RAMC degrades slightly with both front ends, while DH3 improves only with WavLM.

Joint training yields 12.26\% macro DER from the local output and 11.90\% after refinement. The final system improves seven domains over non-joint refinement, chiefly NSF and DH3; AIS worsens from 10.16\% to 10.55\%. Refinement is not uniformly beneficial: on RAMC and DH3, the joint local output slightly outperforms its refined output.

\subsection{Speaker Confusion and Refinement}
The macro components in Table~\ref{tab:main_der} separate identity and detection errors. Frozen-feature refinement reduces SC by 17.4\% with WavLM and 14.6\% with w2v-BERT 2.0. Both reduce MISS but increase FA. For w2v-BERT 2.0, SC falls by 0.44 percentage points, exceeding the 0.16-point increase in FA plus MISS. SC decreases on seven domains; RAMC is the exception. On DH3, SC falls from 2.71\% to 2.51\% and MISS also decreases, but higher FA raises DER overall.

\begin{figure}[!t]
\centering
\includegraphics[width=\linewidth]{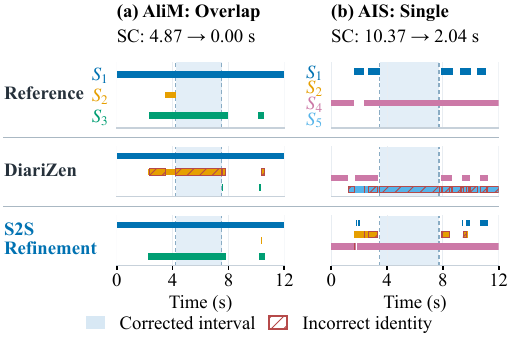}
\caption{Highlighted fixed-count identity corrections with non-joint w2v-BERT 2.0: (a) AliM and (b) AIS.}
\label{fig:sc_cases}
\end{figure}

Figure~\ref{fig:sc_cases} shows count-preserving identity correction: $\{S_1,S_2\}\!\to\!\{S_1,S_3\}$ in AliM overlap and $S_5\!\to\!S_4$ in AIS single-speaker speech. The highlighted intervals last 3.31 and 4.25~s, with zero FA and MISS for both systems. Both corrections select an enrolled speaker that was inactive in the prior at those frames, consistent with the ranking in Eq.~\eqref{eq:prior_fusion}. Figure labels report SC over the full 12-s crops. Both systems use identical full-recording identity mappings, fixed before cropping to exclude local rematching. Residual AIS errors illustrate incomplete correction within a favorable case.

\subsection{Joint-System Gains and Error Trade-offs}
Compared with non-joint refinement, the joint system lowers macro FA and MISS by 0.090 and 0.115 percentage points, while SC rises by 0.024 points. Local SC is also nearly unchanged (2.98\% versus 2.99\%). Detection therefore drives the incremental DER gain.

Within the joint system, refinement still reduces SC from 2.99\% to 2.57\% (13.9\% relative), preserving its identity-correction benefit. Relative to the corresponding local output, the combined FA and MISS increase is 0.05 percentage points with joint refinement, versus 0.16 with non-joint refinement. Thus, despite a slightly smaller SC reduction, joint refinement retains more of this benefit in the final DER.

Joint training lowers final SC on NSF, RAMC, and VoxC but raises it elsewhere. NSF improves all three components; DH3 reduces FA and MISS despite higher SC, while AIS worsens in all three. Lower macro DER therefore does not ensure uniformly better speaker assignment.

With the original refinement parameters, the joint system achieves 11.902\% macro DER, versus 12.077\% for non-joint refinement; retuning changes DER by only 0.006 percentage points. The improvement thus persists without retuning the inference constraints. Initialization, prior sampling, and model selection also differ, so this system comparison does not isolate the joint objective.

\section{Conclusion}
\label{sec:conclusion}
Shared-encoder S2S refinement reuses DiariZen's representations for identity correction without another refinement acoustic encoder. Across eight domains, frozen-feature refinement lowers SC and macro DER at approximately 15\% additional RTF. Joint training with pruned w2v-BERT 2.0 yields the best macro DER, 11.90\%, through complementary detection gains. Component analysis and fixed-count cases demonstrate the value of speaker-conditioned decoding and conservative prior constraints alongside local segmentation.

\clearpage
\section{Acknowledgments}
\noindent\textbf{Generative AI use.} OpenAI Codex (GPT-6) assisted with translating and polishing author-provided text and generating candidate visual elements for clearer presentation in figures and tables. The authors developed all scientific content, manually finalized the manuscript and visuals, verified numerical values, and take full responsibility for the work.

\begingroup
\let\originalthebibliography\thebibliography
\renewcommand{\thebibliography}[1]{\originalthebibliography{#1}\small}

\endgroup
\end{document}